\documentclass[aps,pre,reprint,superscriptaddress,amsmath,amssymb,showpacs,floatfix]{revtex4-1}
\usepackage{graphicx}
\usepackage{color}
\usepackage{CJK}
\usepackage{latexsym}

\begin{document}
\begin{CJK*}{UTF8}{mj}

\title{Block renormalization study on the nonequilibrium chiral Ising
model}

\author{Mina Kim (김민아)}
\affiliation{Department of Physics, University of Seoul, Seoul 130-743,
Korea}
\author{Su-Chan Park (박수찬)}
\affiliation{Department of Physics, The Catholic University of Korea,
Bucheon 420-743, Korea}
\author{Jae Dong Noh (노재동)}
\affiliation{Department of Physics, University of Seoul, Seoul 130-743,
Korea}
\affiliation{School of Physics, Korea Institute for Advanced Study,
Seoul 130-722, Korea}
\date{\today}

\begin{abstract}
We present a numerical study on the ordering dynamics of a
one-dimensional nonequilibrium Ising spin system with chirality. This system is
characterized by a direction-dependent spin update rule.  
Pairs of $+-$ spins can flip to $++$ or $--$ with probability $(1-u)$ or to 
$-+$ with probability $u$ while $-+$ pairs are frozen. 
The system was found to evolve into the ferromagnetic ordered state at
any $u<1$ exhibiting the power-law scaling of the characteristic length scale
$\xi\sim t^{1/z}$ and the domain wall density $\rho\sim
t^{-\delta}$. 
The scaling exponents $z$ and $\delta$ were found to vary continuously with 
the parameter $u$.
In order to establish the anomalous power-law scaling
firmly, we perform the block spin renormalization analysis proposed by Basu
and Hinrichsen [U. Basu and H. Hinrichsen, J. Stat. Mech. (2011) P11023].
Domain walls of $b$ sites are 
coarse-grained into a block spin $\sigma^b$, and the relative frequencies
of two-block patterns $\sigma^b_1 \sigma^b_2$ are measured
in the $b\to\infty$ and $t\to\infty$ limit. These indices are expected to be
universal. By performing extensive Monte Carlo simulations, 
we find that the indices also vary continuously with $u$ and that their
values are consistent with the scaling exponents found in the previous
study. This study serves as another evidence for the claim that the
nonequilibrium chiral Ising model displays the power-law scaling behavior
with continuously varying exponents.
\end{abstract}

\maketitle
\end{CJK*}

\section{Introduction}\label{sec:1}

Macroscopic systems display an intriguing 
dynamic scaling behavior upon ordering~\cite{Glauber63,Bray01}.
When a system in an ordered phase is quenched from a disordered
configuration, the characteristic size $\xi$ of ordered domains increases
with time and microscopic details become less and less important. 
Consequently, there emerges a dynamic scaling
behavior that is classified into a universality class depending on 
symmetry, conservation, and so on. 

Each universality class is characterized by the power-law scaling of the
length scale $\xi \sim t^{1/z}$ with a universal dynamic exponent $z$.
For example, equilibrium systems with a scalar order parameter, such as the
Ising model, have $z=2$ under the nonconserved dynamics and $z=3$ 
under the conserved dynamics in the ordered phase~\cite{Newman90,Bray94}. 
Systems with a vector order parameter also have distinct
values of $z$ depending on the presence of the conservation 
law~\cite{Newman90,Bray94}. 

Recently, the ordering dynamics in a nonequilibrium chiral Ising model~(NCIM) 
was studied numerically in one dimension~\cite{Kim13}. The NCIM, which will
be explained in detail in Sec.~\ref{sec:2}, has two
important features. It has the ferromagnetic states with all spins up or
down as the two equivalent absorbing states. Namely, once the system reaches one of the
two ferromagnetic states, it stays there forever. 
In addition, the NCIM has a direction-dependent 
spin update rule, which makes the system chiral. The chirality breaks 
the spin up-down symmetry.

The model without chirality is equivalent to the nonequilibrium kinetic
Ising model, whose ordering dynamics is described by
$z=2$~\cite{Mussawisade98,Menyhard00}.
When the chirality turns on, 
the dynamic exponent and the other exponents 
are found to vary continuously as a function of a model parameter~\cite{Kim13}.
Such a phenomenon is very rare with only a few
examples~\cite{Lee97,Jain05}.  It might be attributed to the different 
symmetry property of the NCIM. However, its origin is not revealed yet. 
The current status urges us to establish the universality class firmly
by an independent means. 

Basu and Hinrichsen proposed a numerical method to identify a dynamic
universality class by using a block spin
transformation~\cite{Basu11}. Adopting the idea of the real-space
renormalization group transformation~\cite{MK78,K2000}, one divides 
a one dimensional lattice of $L$ sites into $L/b$ blocks of size $b$ 
and coarse-grains 
a spin configuration $\{\sigma_n|n=1,\cdots,L\}$ 
with a block-spin configuration
$\{\sigma^b_n|n=1,\cdots,L/b\}$.
Then, for any pattern 
$c=(xy\cdots)$, 
one can define a correlation
function
\begin{equation}\label{corr_func}
P_c(b,t) = \left\langle 
            \delta(\sigma^b_n(t),x) \delta(\sigma^b_{n+1}(t),y)\cdots
           \right\rangle \ ,
\end{equation}
where $\delta(x,y)$ is the Kronecker delta, 
$\sigma^b_n(t)$ denotes the block spin at site $n$ at time $t$,
and $\left\langle \cdot \right\rangle$ denotes the average over ensembles as
well as $n$. The ratios between the correlation functions
of different patterns turn out to converge to universal values in the 
$t\to\infty$ limit followed by the $b\to\infty$ limit. 
This universal feature was tested for some dynamic universality
classes~\cite{Basu11}.

We apply the block spin analysis to the NCIM in order to confirm that the NCIM
is characterized by the continuously-varying critical 
exponents. In Sec.~\ref{sec:2}, we introduce the NCIM and give a brief
review of the numerical result in Ref.~\cite{Kim13}. Section~\ref{sec:3}
presents the main result of the block spin analysis for the NCIM.
This result is fully consistent with the
previous numerical result and strengthens the claim of the universality
class with the continuously-varying critical exponents. 
The ratio of the correlation functions in Eq.~(\ref{corr_func}) is related
to the critical exponent through a scaling relation. The scaling relation
was proposed in Ref.~\cite{Basu11} on the ground of the scaling ansatz. We
present a microscopic theory for the scaling relation in
Sec.~\ref{sec:4}. We summarize this work with discussions in Sec.~\ref{sec:5}.

\section{Nonequilibrium chiral Ising model}\label{sec:2}

To study a coarsening dynamics of a one dimensional Ising spin chain
$\{s_n=\pm|n=1,\cdots,L\}$ with chirality, 
the authors have suggested the NCIM with the following dynamic rules~\cite{Kim13},
\begin{eqnarray}
+- \stackrel{u}{\longrightarrow} -+ &,\quad&
-+ \stackrel{\bar{u}}{\longrightarrow} +-,\nonumber\\
+- \stackrel{v/2}{\longrightarrow} \begin{cases} ++ \\ -- \end{cases}
&,\quad&
-+ \stackrel{\bar{v}/2}{\longrightarrow} \begin{cases} ++ \\ -- \end{cases} ,
\end{eqnarray}
where $u~(\bar{u})$ and $v~(\bar{v})$ are the transition rates for 
the spin exchange and the single spin flip dynamics of the local configuration
$+-$ ($-+$), respectively. 
We have assumed periodic boundary conditions.
The chirality can be incorporated into the model
by taking different transition rates for $+-$ and $-+$ domain walls.
The NCIM has two equivalent ferromagnetically ordered states 
with all spins up or down. These states are absorbing 
in the sense that the system cannot get out of the states by the above
dynamic rules. 

In addition to its own merit as a minimal model for the chiral dynamics,
the NCIM can be applied to a flocking phenomenon of active Brownian
particles by regarding the Ising spin states $+$ and $-$ as the directions
of motion of particles in one dimension. The flocking model using the active
spins is found in Ref.~\cite{Solon13}.

The chirality breaks the spin up-down symmetry of the Ising model.
Unlike the magnetic field which favors one
of  the spin states, the chirality does not prefer any of the spin 
states. In fact, the NCIM is symmetric under the simultaneous inversion of
spin and space, $s_n \to - s_{L-n+1}$. In higher dimensions, 
this chirality turned out to be irrelevant for Ising-like spin
models with order-disorder transitions~\cite{BS1994} (see also Ref.~\cite{Dutta} for
a generalization to $N$-vector models, which showed that chirality is relevant for $N\ge 2$). 
However the one dimensional system 
with chirality seems to exhibit intriguing scaling behaviors with continuously
varying exponents~\cite{Kim13}. 

It is convenient to map the Ising spin system to a 
reaction diffusion system of two species $A$ and $B$ 
by introducing a random variable $\sigma_n \in \{A,B,O\}$: 
A site $n$ is regarded as being occupied by an $A$ particle~[$\sigma_n = A$] 
if $(s_n s_{n+1})=(+-)$. It is regarded as being occupied by a $B$
particle~[$\sigma_n=B$] if $(s_n s_{n+1})=(-+)$. Otherwise, it is regarded as
being empty~[$\sigma_n=O$].
Within this scheme, all sites are empty in the absorbing states.
Due to the correspondence with Ising spin configurations, 
the two species should be alternating in space
and the number of $A$ particles should be the same as that of $B$ particles.
Under the symmetry operation $s_n \to - s_{L-n+1}$, 
a particle configuration is
mapped to the mirror image with the particle species being invariant.

The spin dynamic rules are translated as follows. With rate $v$ species $A$ hops to one 
of its nearest neighbors chosen with equal probability,
and with rate $u$ species $A$ branches two $A$'s at both nearest neighbor 
sites and it changes to another species~($A \to ABA$). 
The dynamics of species $B$ is the same as above with rates given by the 
barred parameters. Whenever two species happen to
occupy the same site by either hopping or branching event, both particles annihilate 
immediately~($A+B \to O$).

\begin{figure*}[t]
\includegraphics*[scale=1.3]{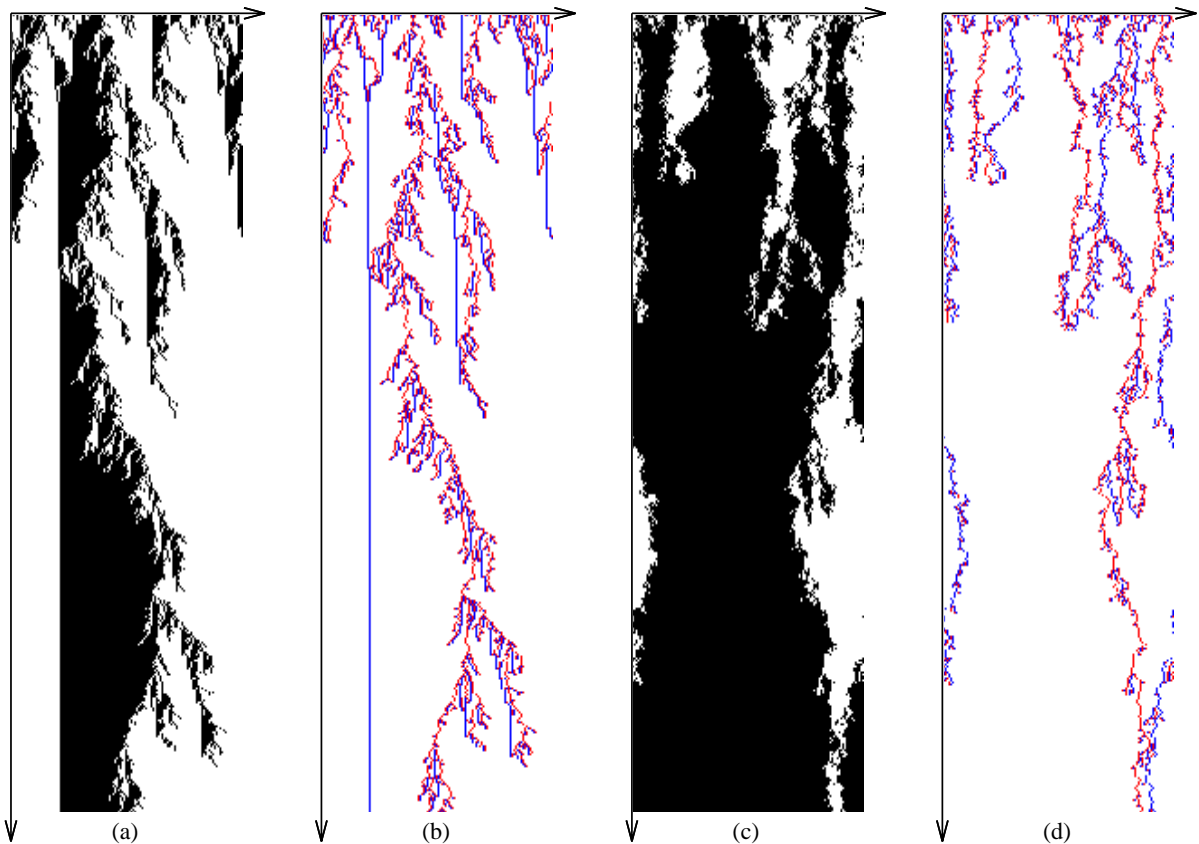}
\caption{\label{fig1} (Color online) Space-time patterns of the Ising spins
with chirality~($\bar{u}=\bar{v}=0$ and $u=1-v=0.5$) in (a) and (b), and
without chirality~($u=v=\bar{u}=\bar{v}=0.5$) in (c) and (d). 
Spin dynamics are shown in (a) and (c), where black~(white) pixels
represent sites of $+~(-)$ states. Particle dynamics are shown in (b) and (d),
where red~(blue) pixels represent $A~(B)$ particle.
The horizontal and vertical directions correspond to the
spatial and temporal directions, respectively.}
\end{figure*}

Time evolution of the NCIM with $\bar{u}=\bar{v}=0$ is illustrated in terms of the spin variable 
$\{s_n\}$ in Fig.~\ref{fig1}(a) and in terms of the particle
variable $\{\sigma_n\}$ in Fig.~\ref{fig1}(b). 
The chirality gives rise to an interesting space-time pattern.
The motions of $A$ and $B$ species are asymmetric, while
none of the spin states are preferred. As the dynamics proceeds, 
a characteristic domain size increases and the density of particles
decreases with time.  
The ordering or coarsening dynamics is characterized by the power-law scaling
of the characteristic domain size  
\begin{equation}\label{z_def}
\xi(t) \sim t^{1/z} \ , 
\end{equation}
and the domain wall or particle density
\begin{equation}\label{delta_def}
\rho(t) \sim t^{-\delta}, 
\end{equation}
with the dynamic exponent $z$ and the density decay exponent $\delta$.

Without chirality~($u=\bar{u}$ and $v=\bar{v}$), the NCIM reduces to the
nonequilibrium kinetic Ising model~(NKIM) or the reaction diffusion 
system~\cite{Mussawisade98,Menyhard00,benAvraham94,Takayasu92}~(see 
Figs.~\ref{fig1}(c) and(d)) which
is exactly solvable~\cite{benAvraham94}. The exact solution
reveals that 
the ordering dynamics belongs to the universality class of the model at
$u=\bar{u}=0$. It corresponds to the Ising model under the zero-temperature
Glauber dynamics~\cite{Glauber63}, or equivalently the voter
model~\cite{Cox89,Ligget95}. The critical exponents are $z=2$ and $\delta=1/2$.

When the chirality sets in~($u\neq \bar{u}$ or $v\neq \bar{v}$), the model
is not solvable any more. 
The model has been studied in various regions of the
parameter space. For example, 
when $v=\bar{v}$ and $u\neq \bar{u}$, it becomes the
mixture of the asymmetric simple exclusion process and the voter
model studied in Ref.~\cite{Belitsky01,MacPhee10}.
The ordering dynamics of the NCIM has been studied 
numerically in Ref.~\cite{Kim13}. 
Surprisingly, the numerical study reveals that
the dynamic exponent and the density decay exponent vary continuously 
within the range $1< z \leq 2$ and $0< \delta \leq 1/2$.
We will provide an independent evidence for the continuously varying critical 
exponents in the following sections.

\section{Block spin analysis}\label{sec:3}
At criticality, the scaling functions as
well as the critical exponents are universal. Extending this idea, Basu and
Hinrichsen~\cite{Basu11} proposed that the spatial correlation of spins
in the long time and large distance limit can be used 
in identifying a dynamic universality class. This is accomplished by
coarse-graining a `spin' configuration with that of a `block spin'.
As in the real-space renormalization group transformation~\cite{MK78, K2000}, 
$b$ spins in a row are coarse-grained by a single block spin.  
Then, large-distance correlations are measured in terms of the block
spins in the $b\to\infty$ limit.

We apply the coarse graining scheme to the particle or domain wall 
variable $\sigma_n \in \{A,B,O\}$ of the NCIM. 
The coarse-graining should preserve the symmetry and the conservation of the
system.
It should also preserve the absorbing nature of the vacuum state.
The following coarse-graining scheme fulfills the requirements. 

To a given block of size $b$, the number of $A$ and $B$ particles are 
denoted by $N(A)$ and $N(B)$, respectively. 
If $N(A)=N(B)=0$, the block is in a vacuum state and 
it is assigned to a state ${O}$. 
If $N(A)>N(B)$, the block separates the
$+$ domain in the left from the $-$ domain in the right. 
Thus it is assigned to a state $A$.
If $N(A)<N(B)$, the block separates the
$-$ domain in the left from the $+$ domain in the right, so 
it is assigned to a state $B$. If $N(A)=N(B)\neq 0$, the block is not
in the vacuum state, nor does it separate different domains. Hence we need
to assign a block state different from $A$, $B$, and $O$. 
Furthermore, due to the chirality, we need to assign a different block 
state depending on
whether the domain walls have an $AB$ ordering or $BA$ ordering. We will
assign a block state $X$ for the former case and $Y$ for the latter. The
coarse-graining rule is summarized below:
\begin{equation}
\label{5-state_mapping}
\sigma^b = 
\begin{cases}
{O} & \mbox{if } N(A)=N(B)=0 \\
{A} & \mbox{if } N(A)>N(B)  \\
{B} & \mbox{if } N(A)<N(B)  \\
{X} & \mbox{if } N(A)=N(B)\neq 0 \mbox{ and $AB$ ordering} \\
{Y} & \mbox{if } N(A)=N(B)\neq 0 \mbox{ and $BA$ ordering}
\end{cases} \ .
\end{equation}

Note that the block spin $\sigma^b$ takes on five different states.
This is in contrast to the Ising system without chirality
where one needs only three different block states~\cite{Basu11}. 
Due to the chirality, ${ A}$ and ${ B}$
should be distinguished, so should ${ X}$ and ${ Y}$.
Under the symmetry operation $s_n \to - s_{L-n+1}$, ${A}$ and ${B}$
remain the same while ${X}$ is transformed to ${Y}$ and vice versa.

Using the coarse-graining rule, we evaluate numerically 
the correlation function
$P_c(b)$ defined in Eq.~(\ref{corr_func}) especially for all two-blocks
patterns 
\begin{equation}\label{c_pattern}
\begin{split}
 c \in & \{ {OO},{OA},{AO},{OB},{BO},{AB},{BA},
            {XO},{OX}\\
& 
{YO},{OY},{XA},{BX},{AY},{YB},{XX},{YY} \}.
\end{split}
\end{equation}
Patterns ${AA}$, ${BB}$, ${XB}$, ${AX}$,
${BY}$, ${YA}$, ${XY}$, and ${YX}$ are forbidden by the background spin dynamics. 
We concentrate on the model with $\bar{u}=\bar{v}=0$ and $u+v=1$, 
which was referred to as the maximum chiral model (MCM) in Ref.~\cite{Kim13}.
In this model, $A$ particles branch with the probability $u$ and hops with
the probability $v=1-u$ while 
$B$ particles are frozen except when the instantaneous
pair annihilation~($A+B\to O$) occurs. 

Monte Carlo simulations are performed in systems of sizes
$L=2^{24}$ at $u=0.0$ and $0.1$, $L=2^{23}$ at $u=0.2$ and $0.3$, 
$L=2^{22}$ at $u=0.4$, $0.5$, and $0.6$,
$L=2^{21}$ at $u=0.7$, $0.8$, and $0.9$, and
$L=2^{20}$ at $u=1.0$. 
The initial configuration is taken to be the fully occupied state $(\cdots
ABAB\cdots)$ that is equivalent to the antiferromagnetic state 
$(\cdots +-+-\cdots)$. 
During simulation, the correlation functions $P_c(b,t)$ are evaluated
for all two-blocks patterns $c$ in Eq.~(\ref{c_pattern}) 
at times $t=2^l$ with $l\leq 24$.
The block sizes are $b=2^{k}$ with $k\leq 5$.
All the data are obtained by averaging over $N_{S}\le 5000$ independent 
samples. 

\begin{figure}[t]
\includegraphics*[width=\columnwidth]{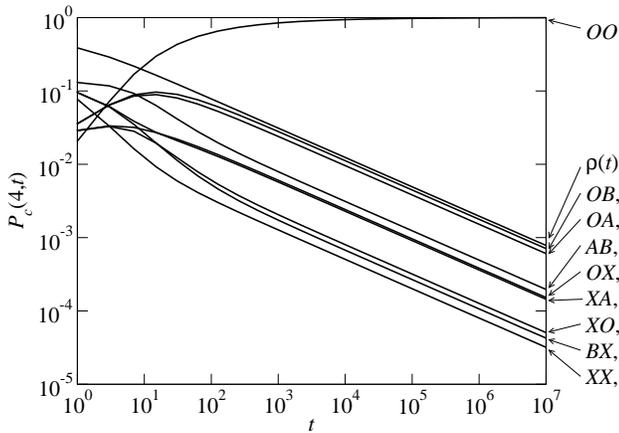}
\caption{\label{fig2} Temporal decay of the two-blocks 
correlation function $P_{c}(b=4)$ at the model parameter $u=0.3$. Also shown
is the overall particle density $\rho(t)$. 
}
\end{figure}

Figure~\ref{fig2} presents the two-blocks correlation functions with $b=4$ in
the MCM with $u=0.3$. After a transient period, all 
the correlation functions except for the pattern $c={OO}$ decay
algebraically with the density decay exponent $\delta$. Since the
system eventually orders, $P_{{OO}}$
converges to $1$ in the $t\to\infty$ limit. 
This temporal scaling is also observed for other values of $b$:
\begin{equation}
\label{power-law}
P_{c} (b,t) \sim t^{-\delta} \quad \text{for} \quad c \ne {OO}\ .
\end{equation}
Note that the correlation functions are not independent of each other. 
The symmetry under $s_n \to - s_{L-n+1}$ requires
that  
\begin{equation}
\begin{split}
\label{symmetry}
P_{{OA}}=P_{{AO}},\quad P_{{OB}}=P_{{BO}},\\
P_{{AB}}=P_{{BA}},\quad P_{{XO}}=P_{{OY}},\\
P_{{OX}}=P_{{YO}},\quad P_{{XX}}=P_{{YY}},\\
P_{{XA}}=P_{{AY}},\quad P_{{BX}}=P_{{YB}}.
\end{split}
\end{equation}

Following Ref.~\cite{Basu11}, we define
\begin{equation}
S_{c}(b,t)\equiv \frac{P_{c}(b,t)}{\sum_{c' \ne {OO}}{P_{c'}(b,t)}}
=\frac{P_{c}(b,t)}{1-P_{{OO}}{(b,t)}}.
\end{equation}
It measures the relative frequency of a block pattern $c$ among 
all patterns but the vacuum pattern ${OO}$. Upon taking the ratio, the
temporal dependence cancels out and the amplitudes determine $S_c(b,t)$.
The scale invariance
suggests that the quantity should converge to a universal
value~\cite{Basu11} 
\begin{equation}
S_c = \lim_{b\to\infty} S_c(b)
\end{equation}
with
\begin{equation}
S_c(b) = \lim_{t\to\infty} S_c(b,t)  \ .
\end{equation}

\begin{figure}[t]
\includegraphics*[width=\columnwidth]{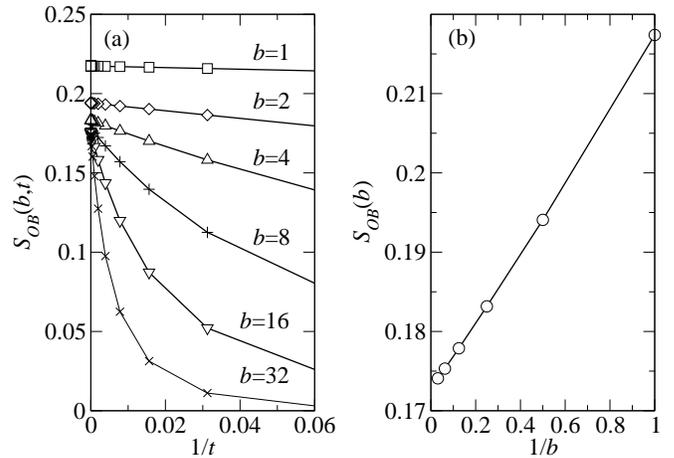}
\caption{\label{fig3} (a) $S_{{OB}}(b,t)$ for $b=1, 2, \ldots, 32$ (from top to
bottom) and (b) $S_{{OB}}(b)$ at the model parameter $u=0.3$.}
\end{figure}

Figure~\ref{fig3}(a) presents the relative frequency $S_{{OB}}(b,t)$ 
of a pattern ${OB}$ at several levels of coarse graining at $u=0.3$. 
It converges to a constant value $S_{{OB}}(b)$
in the $t\to\infty$ limit. The extrapolated values are plotted as a function
of $1/b$ in Fig.~\ref{fig3}(b), from which we can estimate $S_{{OB}}$.
In practice, we adopted a power-law fitting to the forms
$S_c(b,t) =  S_{c}(b) + a t^{-\chi}$ and $S_{c}(b) = S_{c} + a' b^{-\chi'}$. 

\begin{figure}[t]
\includegraphics*[width=\columnwidth]{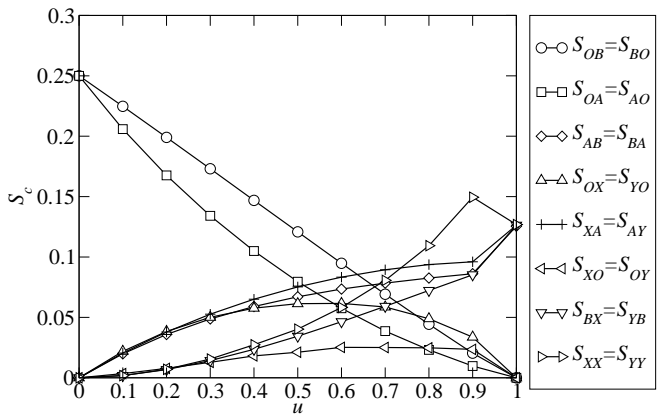}
\caption{\label{fig4} $S_c$ for two-blocks patterns $c$.
Lines are a guide to an eye. }
\end{figure}

Repeating the same procedure, we obtain the relative frequency for all
patterns. They are presented in Fig.~\ref{fig4}.
At $u=0$, $A$ particles diffuse without branching and annihilate in pairs 
with $B$ particles upon collision. Hence, when $t\gg b^2$, the block spin
configurations consist of isolated ${A}$s and ${B}$s in the sea of
${O}$s. It explains the numerical result that 
$S_{{OB}}=S_{{BO}}=S_{{OA}}=S_{{AO}}=1/4$ with the others being zero.
At $u=1$, the system reaches an active steady state with a finite particle 
density. Block spins are in a state of ${A}$, ${B}$, ${X}$,
or ${Y}$ equally likely, and a spatial correlation is absent in the
$b\to\infty$ limit. Thus, $S_{{AB}}=S_{{BA}}=S_{{XA}}=S_{{AY}}=
S_{{BX}}=S_{{YB}}= S_{{XX}}=S_{{YY}}=1/8$ and all the others 
are zero. 

As is noticeable in Fig.~\ref{fig4}, $S_c$ seems discontinuous at $u=1$.
The model at $u=1$ is a singular limit 
in the sense that there is no chance of falling into absorbing
states once initial particle density is finite. Thus unlike the case
of $u<1$, $P_{OO}(b,t)$ cannot approach 1 with $t$.
We speculate that the sharp change of $S_c$'s near $u=1$ could be caused 
because $P_{OO}(b,\infty)$ changes abruptly at $u=1$.

As the model parameter $u$ varies, each value of $S_{c}$ varies continuously.
Under the hypothesis that $S_c$ should be universal, 
Fig.~\ref{fig4} provides an evidence for the continuously
varying critical exponents of the NCIM. 
In Sec.~\ref{sec:4}, we will estimate the critical exponent $z\delta$
using the values of $S_c$'s.

\section{\label{sec:4}Critical exponent $z\delta$}
For general one-dimensional models with absorbing states, 
one can introduce a random variable $\rho_n$ at each site $n$ ($n=1,2,\ldots,L$) 
which takes either 1 or 0.
Conventionally, $\rho_n$ is defined in such a way that a configuration 
is absorbing if and only if $\rho_n = 0$ for all $n$. In this section, 
however, we only assume that the condition $\rho_n = 0$ for all 
$n$ is a necessary condition of system's being in one of absorbing states.
Still, however, the average of $\rho_n$ over space and ensemble 
\begin{equation}
\rho(t) = \frac{1}{L} \sum_n \langle \rho_n \rangle
\end{equation} 
can play the role of an order parameter. 
{\em For convenience}, we will say that a site
$n$ is occupied (vacant) if $\rho_n = 1$ (0), even though
$\rho_n = 0$ does not necessarily imply that the site $n$ is truly devoid
of any particles of the background dynamic model.
If we limit ourselves to the stochastic behavior of $\rho_n$ instead of the
domain wall variables $\sigma_n$, the block configurations become simpler
than those in Sec.~\ref{sec:3}. A block of size $b$ is assigned to be occupied
only when it contains at least one occupied site. 
The specific choice of $\rho_n$ for the NCIM will be taken later.

Combining Eqs.~(\ref{z_def}) and (\ref{delta_def}), the density scales as
\begin{equation}
\rho \sim \xi^{-\alpha}
\end{equation}
with the exponent
\begin{equation}
\alpha = z \delta .
\end{equation}
Under the assumption of the scale invariance during the critical dynamics, 
it is claimed in Ref.~\cite{Basu11} that
\begin{equation}\label{zd}
\lim_{b\to\infty} \lim_{t\to\infty} \frac{ P_1(b,t) }{ P_1(2b,t)}
 = 2^{-\alpha }  ,
\end{equation}
where $P_1(b,t)$ is the probability that a block of size $b$ is occupied,
that is, it contains at least one occupied site. 
Formally speaking, $P_1(b,t)$ is defined as
\begin{equation}
P_1(b,t) = \lim_{L\rightarrow \infty} \frac{1}{L} \sum_n \left \langle 1 - V_{n,b} \right \rangle,
\label{rho_cg}
\end{equation} 
where $V_{n,b} \equiv \prod_{r=0}^{b-1} (1-\rho_{n+r})$ 
with $V_{n,0} \equiv 1$.
Note that $(1-V_{n,b})$ can be interpreted as the `block spin' in the sense of 
Ref.~\cite{Basu11}.
We will provide a general microscopic theory for the condition under which 
the relation~\eqref{zd} is valid.

To analyze Eq.~\eqref{zd} systematically,
we introduce three types of correlation functions such as 
\begin{align}
P_{\rho\rho}(r,t) &= \frac{1}{L} \sum_n \left \langle \rho_n \rho_{n+r} \right \rangle,\\
P_{\rho v\rho}(r,t) &= \frac{1}{L} \sum_n \left \langle \rho_n  V_{n+1,r-1}\rho_{n+r} \right \rangle,\\
P_{v\rho}(r,t) &= \frac{1}{L} \sum_n \left \langle  V_{n,r}\rho_{n+r}  \right \rangle 
.
\label{Eq:Pvr}
\end{align}
Taking the translational invariance for granted, 
$P_{\rho\rho}(r,t)$ is the joint probability that two sites separated
by a distance $r$ are occupied simultaneously. Similarly, $P_{\rho
v\rho}(r,t)$ denotes the joint probability that two sites separated by a
distance $r$ are occupied with all intermediate sites being vacant.
$P_{v\rho}(r,t)$ is the joint probability that a site is occupied and
preceded by $r$ empty sites.
For example, $P_{\rho v\rho}(1,t) = \langle \bullet \bullet \rangle$,
$P_{v\rho}(1,t) = \langle \circ \bullet\rangle$,
$P_{\rho v\rho}(2,t) = \langle \bullet \circ \bullet \rangle$,
$P_{ v\rho}(2,t) = \langle \circ \circ \bullet \rangle$, and so on, where
$\bullet$ ($\circ$) signifies an occupied (a vacant) site.

The first step is to represent $\rho(t)$ and $P_1(b,t)$
in terms of these correlation functions. 
The identity $V_{n,1}+\rho_n=1~(\circ+\bullet=1)$ yields that
$\rho(t) = P_{v\rho}(1,t)+ P_{\rho v \rho}(1,t)~(\langle \bullet\rangle =
\langle \circ\bullet \rangle + \langle \bullet \bullet \rangle)$ and 
$P_{v\rho}(r-1,t)=P_{\rho v \rho}(r,t)+ P_{v\rho}(r,t)~(\langle \circ
\cdots\bullet\rangle = \langle \bullet \circ \cdots\bullet \rangle + \langle
\circ\circ\cdots\bullet\rangle)$. Applying the second relation iteratively,
we get, for any $1 \le b \le L$, 
\begin{equation}
\rho(t) = \sum_{r=1}^b P_{\rho v \rho}(r,t) + P_{v\rho}(b,t).
\label{Eq:rhoPP}
\end{equation}
In the following discussion, $L \rightarrow \infty$ limit is assumed 
to be taken first.
Note that under the thermodynamic limit $\rho(t) > 0$ for finite $t$ once 
$\rho(t=0) > 0$ and no sample can fall into one of absorbing states 
up to finite $t$.

Using the identity $V_{n,1}=1-\rho_n$ again, one can
decompose $V_{n,b} = V_{n,b-1} V_{n+b-1,1}$ into $V_{n,b}= V_{n,b-1} -
V_{n,b-1}\rho_{n+b-1}$. Hence, we obtain $P_1(b,t) = P_1(b-1,t) +
P_{v\rho}(b-1,t)$. Applying the relation iteratively and using $P_1(1,t) =
\rho(t)$, we can rewrite $P_1(b,t)$ as
\begin{align}
P_1(b,t)
&= \rho(t) + \sum_{r=1}^{b-1} P_{v\rho}(r,t)\nonumber\\
&= \rho(t) + \sum_{r=1}^{b-1} \left ( \rho(t) - \sum_{k=1}^r P_{\rho v\rho}(k,t) \right )\nonumber \\
&= b \rho(t) - \sum_{r=1}^{b} ( b - r ) P_{\rho v\rho}(r,t),
\end{align}
where the relation (\ref{Eq:rhoPP}) is used in the second line.
Consequently we obtain 
\begin{align}
R(b,t) \equiv \frac{P_1(b,t)}{P_1(2b,t)}
=\frac{\displaystyle b - \sum_{r=1}^{b} (b - r) F(r,t)}
{\displaystyle 2b - \sum_{r=1}^{2b} (2b - r) F(r,t)},
\label{Eq:Rb}
\end{align}
where 
\begin{equation}\label{F_def}
F(r,t) \equiv P_{\rho v \rho}(r,t)/\rho(t) .
\end{equation}
It can be interpreted as the conditional probability of 
$V_{n+1,r-1}\rho_{n+r}= 1$ given that $\rho_n =1$. 
Namely, $F(r,t)$ is the probability that a given particle would
find its first neighbor particle at distance $r$ and at time $t$.

From Eq.~\eqref{Eq:rhoPP}, we find a normalization condition
\begin{equation}
\sum_{r=1}^b F(r,t) + \frac{P_{v\rho}(b,t)}{\rho(t)}=1.
\end{equation}
According to the probability interpretation of $F(r,t)$ above, 
we can claim that
\begin{equation}
\sum_{r=1}^\infty F(r,t) \equiv \lim_{b\to\infty}\sum_{r=1}^b F(r,t) = 1,
\label{Eq:F_blimit}
\end{equation}
which is equivalent to
\begin{equation}\label{Eq:norm2_ft}
\lim_{b\rightarrow \infty}  \frac{P_{v\rho}(b,t)}{\rho(t)}
=0
\end{equation}
for any $t$. Since $\rho(t)$ is finite for finite $t$, 
Eq.~\eqref{Eq:norm2_ft} should be satisfied because the mean distance between
two occupied sites should be $1/\rho(t)$. Recall that
the thermodynamic limit is assumed to be taken already. 

It is quite tempting to claim that
\begin{equation}
\lim_{b\rightarrow \infty} \lim_{t\rightarrow \infty} \frac{P_{v\rho}(b,t)}{\rho(t)}
=0
\label{Eq:norm_cond}
\end{equation}
and 
\begin{equation}
\sum_{r=1}^\infty F_\infty(r) = 1,
\label{Eq:norm}
\end{equation}
where $F_\infty(r) = \lim_{t\to \infty} F(r,t)$. 
Unfortunately, however, this is not always true. 
A counter example can be found from the pair annihilation 
model~($A+A \rightarrow 0$). 
In this example, we define $\rho_n$ such that $\rho_n=1$ if a particle is 
present at site $n$ and 0 otherwise.
Since $\rho(t) \sim t^{-1/2}$ and $P_{\rho\rho}(r,t) \sim r
t^{-3/2}$~\cite{DZ1996,BM1999,PPK2001,MB2001}
we get
\begin{equation}
0 \le F(r,t) \le P_{\rho\rho}(r,t)/\rho(t) \sim t^{-1},
\end{equation}
where we have used $P_{\rho v\rho}(r,t) \le P_{\rho\rho}(r,t)$.
Thus, $F_\infty(r) = 0$ for all $r$, which cannot be consistent with 
Eq.~\eqref{Eq:norm}.

The normalization condition Eq.~\eqref{Eq:norm} for $F_\infty(r)$ is not satisfied when vacant sites form an 
infinite interval in the $t\to\infty$ limit. Therefore, we introduce a
parameter $0\le \phi \le 1$ such that
\begin{equation}
\sum_{r=1}^\infty F_\infty(r)  =  1 - \phi.
\label{Eq:norm_phi}
\end{equation}
Then, the numerator of Eq.~\eqref{Eq:Rb} can be written as
\begin{align}
G(b) &\equiv b - \sum_{r=1}^{b}(b-r) F_\infty(r)\nonumber\\
&= b\phi + b \sum_{r=b+1}^\infty F_\infty(r) + \sum_{r=1}^{b} r F_\infty(r).
\end{align}
Assuming the scale invariance, we expect
$F_\infty(r) \sim r^{-\theta}$ with a critical exponent $\theta$ which should
be larger than 1 by Eq.~\eqref{Eq:norm_phi}. Within this assumption, one
can easily see that
\begin{equation}
b \sum_{r=b+1}^\infty F_\infty(r) \sim \sum_{r=1}^b r F_\infty(r) \sim 
b^{\min[0,2-\theta]}
\ll b
\end{equation}
for large $b$. 

Suppose that $\phi$ is strictly positive. 
Then, $G(b) \simeq bc + O(b^{\min[0,2-\theta]})$ and 
\begin{equation}
\lim_{t\to\infty}R(b,t) = 2^{-1} \ ,
\label{Eq:fortuitous}
\end{equation}
which gives $\alpha=1$. 
The pair annihilation model belongs to this category with $\phi=1$. Since the
model is characterized with $z=2$ and $\delta=1/2$, 
the relation \eqref{zd} appears to be valid. 
However, we believe that this coincidence is fortuitous. 
As a counter example, consider the two-species 
diffusion-limited annihilation model~($A+B \rightarrow 0$) 
and interpret $\rho_n$ as the particle
occupation number irrespective of species. If the system evolves from 
a random initial condition, $\rho(t) \sim t^{-1/{2z}}$ with
$z=2$~\cite{Toussaint83,Kang84,Bramson88,Leyvraz92}. 
Since inter-particle distances diverges as $1/\rho \sim t^{1/2z}$ and
there is no branching event which can place a particle close to a given 
particle, $F_\infty(r)$ should be 0 for finite $r$.
Thus, Eq.~\eqref{zd} leads to $\alpha=1$
that is different from $z\delta = 1/2$.
In other words, unlike the general idea of the renormalization group, Eq.~\eqref{zd} 
has limited applicability when the normalization in Eq.~\eqref{Eq:norm} fails.

If the normalization is valid ($\phi=0$) and 
$F_\infty(r) \sim r^{-\theta}$ for large $r$, 
the asymptotic behavior of $G(b)$ becomes
\begin{equation}
G(b) \sim \begin{cases} b^{2-\theta} & 1< \theta<2,\\
\ln b & \theta = 2,\\
\text{const} & \theta>2,
\end{cases}
\end{equation}
which results in
\begin{equation}\label{Eq:limR}
\lim_{b\rightarrow\infty} \lim_{t\to\infty}
R(b,t) = \begin{cases} 2^{-(2-\theta)}, & 1 < \theta < 2, \\
1, & \theta \ge 2.
\end{cases}
\end{equation}

Assuming that Eq.~\eqref{zd} is valid with $\alpha = z\delta$ for any $\theta$,
Eq.~\eqref{Eq:limR} suggests that $\alpha$ should be zero for $\theta\ge 2$.
Since $z$ cannot be zero, $\delta$ should be zero if $\theta \ge 2$.
That is, the system with $\theta \ge 2$ should be in the active phase and $F_\infty(t)$ 
should actually decay exponentially. 
Thus, the necessary conditions that a {\em critical} system satisfies Eq.~\eqref{zd} are
Eq.~\eqref{Eq:norm} and $F_\infty(r) \sim r^{-\theta}$ with $1 < \theta < 2$ 
(or $ \alpha < 1$) for sufficiently large $r$.

Assuming that all necessary conditions are satisfied, we will now argue that 
$2 - \theta$ is indeed equal to $\alpha$. We start from the
observation that
\begin{align}
\langle \rho_i \rho_{i+r} \rangle
=&P_{\rho v \rho}(r,t)+\sum_{k=1}^{r-1} \left \langle \rho_i V_{i+1,k-1}
\rho_{i+k} \rho_{i+r} \right \rangle , 
\end{align}
where we have exploited the translational invariance of the system.
Employing a cluster mean field-type approximation such that
\begin{equation}
\left \langle \rho_n V_{i+1,k-1}
\rho_{n+k} \rho_{n+r} \right \rangle \approx
\frac{P_{\rho v \rho}(k,t)P_{\rho\rho}(r-k,t)}{\rho(t)},
\label{Eq:CMF}
\end{equation}
we get
\begin{equation}
C_\infty(r) \approx F_\infty(r) + \sum_{k=1}^{r-1} F_\infty(k) C_\infty(r-k),
\end{equation}
where $C_\infty(r) = \lim_{t\to\infty} P_{\rho\rho}(r,t)/\rho(t)$.
Introducing generating functions 
\begin{equation}
\widetilde C_\infty(s) = \sum_{r=1}^\infty e^{-s r} C_\infty(r),\quad
\widetilde F_\infty(s) = \sum_{r=1}^\infty e^{-s r} F_\infty(r),
\end{equation}
and using the convolution theorem, we get
\begin{equation}
\widetilde C_\infty(s) \approx \frac{\widetilde F_\infty(s)}{1 - \widetilde
F_\infty(s)}.
\label{Eq:CF}
\end{equation}

\begin{figure}[t]
\includegraphics*[width=\columnwidth]{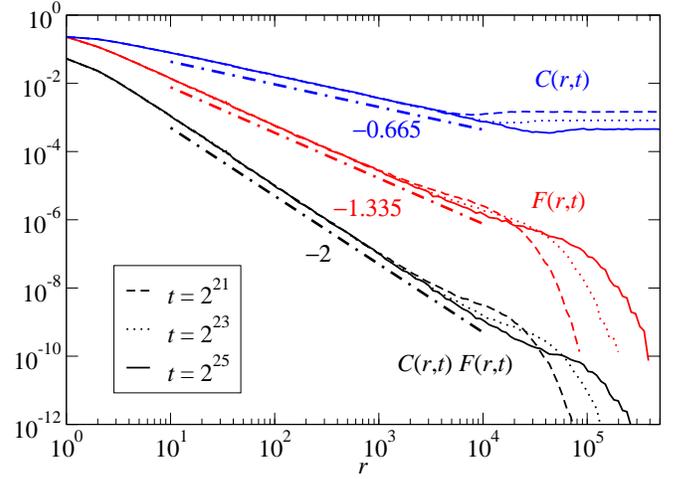}
\caption{(Color online) Correlation functions $C(r,t)$~(blue) and
$F(r,t)$~(red), and their product~(black). The dot-dashed lines with
indicated slopes are guides to the eyes. The data are obtained for the
NCIM with the parameter values $u=1-v=0.3$ and $\bar{u}=\bar{v}=0$. The
lattice size is $L=2^{20}$ and the simulation time is up to $t=2^{25}$. 
The data are averaged over more than 100 samples and log-binned.}
\label{fig5}
\end{figure}
From the scale invariance, we expect $C(r) \sim r^{-\alpha}$ and, in turn,
\begin{align}
\widetilde C_\infty(s) &= \sum_{r=1} C_\infty(r) e^{-sr} \approx 
\int_1^\infty r^{- \alpha} e^{-sr}dr \nonumber\\
&\sim s^{-(1- \alpha)} \int_0^\infty u^{-\alpha} e^{-u} du
\sim s^{-(1-\alpha)},
\end{align}
which diverges as $s \rightarrow 0$ if $ \alpha < 1$ (recall that this 
is one of necessary conditions).
Since $\widetilde F_\infty(s) \rightarrow 1$ as $s \rightarrow 0$ due to
Eq.~\eqref{Eq:norm}, $1 - \widetilde F_\infty(s) $ should approach  
0 as $s \to 0$ for Eq.~\eqref{Eq:CF} to be valid. 
For small $s$, we obtain
\begin{align}
1 - \widetilde F_\infty(s) &= \sum_{r=1}^\infty F_\infty(r) ( 1 - e^{-sr})
\approx \int_1^\infty r^{-\theta} ( 1 - e^{-sr} ) dr\nonumber\\
&= s^{\theta - 1}\int_s^\infty u^{-\theta} ( 1 - e^{-u} ) du .
\end{align}
When $1 < \theta < 2$, the integral part converges to a finite constant as
$s\to 0$, so $1-\widetilde F_\infty(s) \sim s^{\theta - 1}$. Plugging this
into Eq.~\eqref{Eq:CF}, we obtain the scaling relation
\begin{equation}\label{theta_alpha}
\theta = 2 - \alpha \ .
\end{equation}
If we use the scaling relation in Eq.~\eqref{Eq:limR}, 
we finally arrive at the relation~\eqref{zd} with $\alpha=z \delta$.

The scaling relation~\eqref{theta_alpha} is tested
numerically for the NCIM. We measured the correlation functions $F(r,t)$ and
$C(r,t)$ numerically in Monte Carlo simulations.
Figure~\ref{fig5} presents the numerical data for the system of size
$L=2^{20}$ with the parameters $u=1-v=0.3$ and $\bar{u}=\bar{v}=0$.
The correlation functions follow a power law in the regime $r\ll t^{1/z}
\ll L$. The power law justifies the requirement for the scaling argument. 
We also plot the product of $C(r,t)$ and $F(r,t)$. It follows the power law
with the exponent $-2$, which verifies the scaling
relation~\eqref{theta_alpha}. The same results are obtained from other
values of $u$ (details not shown here).
Thus, we expect that the cluster man-field approximation
leads to the correct scaling relation.

The remaining question is why the cluster mean-field type approximation
should be accurate even though the fluctuation is crucial in one dimension.
The cluster mean-field approximation has the same spirit as the independent
interval approximation~\cite{Alemany95,Krapivsky97} which 
was successful to describe the domain size distribution in reaction 
diffusion systems. Of course, a successful approximation in one model does not necessarily 
imply the applicability to any other models. It can be an interesting theoretical challenge
to understand the applicability of the cluster mean-field type approximation Eq.~\eqref{Eq:CMF}
which is beyond the scope of this work. We defer this question to later works.

\begin{figure}[t]
\includegraphics*[width=\columnwidth]{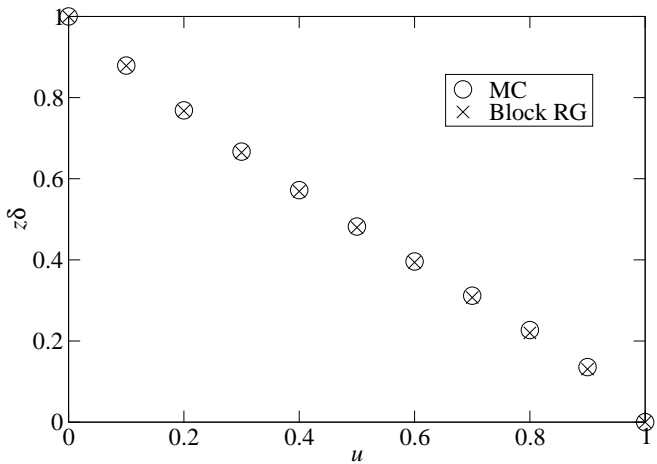}
\caption{\label{fig6} Comparison of the scaling exponent $z\delta$ obtained
from the Monte Carlo simulation of Ref.~\cite{Kim13} 
and the block spin analysis.}
\end{figure}
Accepting the relation~\eqref{zd}, we estimate the critical exponent 
$\alpha = z\delta$ of the NCIM using the indices $S_c$ measured 
in the previous section. 
First we need to define the random variable $\rho_n$. 
We set $\rho_n = 1$ if site $n$ is occupied by a particle
irrespective of its species and 0 if site $n$ is empty.
With this definition, $P_1(b,t)=1-P_O(b,t)$ becomes 
\begin{equation}
P_1(b,t) = P_A (b,t)+P_B(b,t)+P_X(b,t)+P_Y(b,t) \ .
\end{equation}
It is convenient to rewrite $P_1(b,t)$ in terms of two-blocks correlation
functions.
A block of $\rho_n=A$ may be followed by a block of  $\rho_{n+1}=O$, $B$, or
$Y$. It yields that $P_A(b,t) = P_{AO}(b,t)+P_{AB}(b,t)+P_{AY}(b,t)$. 
One can find the corresponding relations for the others. Thus, we have
\begin{eqnarray*}
P_1(b,t) &=& P_{AO}(b,t)+P_{AB}(b,t)+P_{AY}(b,t) \\
         &+& P_{BO}(b,t)+P_{BA}(b,t)+P_{BX}(b,t) \\
         &+& P_{XO}(b,t)+P_{XA}(b,t)+P_{XX}(b,t) \\
         &+& P_{YO}(b,t)+P_{YB}(b,t)+P_{YY}(b,t)  \ .
\end{eqnarray*}
Dividing this with $P_1(2b,t) = 1- P_{OO}(b)$ and taking the limits, we 
obtain 
\begin{align}\label{zd_ratio}
2^{-\alpha} =& S_{AO}+S_{AB} + S_{AY} + S_{BO} + S_{BA} + S_{BX} +\\& S_{XO} + 
S_{XA}+S_{XX} + S_{YO} + S_{YB} + S_{YY}.
\nonumber
\end{align}

We evaluate the critical exponent $\alpha=z\delta$ by inserting the
numerical values of $S_c$'s into Eq.~\eqref{zd_ratio}. For example, we
obtain that $\alpha \simeq 0.665$ at $u=0.3$. This value is in perfect
agreement with the power-law decay of the correlation function $C(r,t)\sim
r^{-\alpha}$ in Fig.~\ref{fig5}.
It is also consistent with the power-law scaling of $F(r,t)\sim r^{-\theta}$
with $\theta=2-\alpha$.
In Fig.~\ref{fig6}, the numerical results for $\alpha$, thus obtained, are compared  
with the values obtained from Monte Carlo simulations in Ref.~\cite{Kim13}. 
Both data are in excellent agreement with each other.

We also studied the scaling relation by assigning $\rho_n=1$ if site $n$ is
occupied by $A$ and $\rho_n=0$ if site $n$ is occupied by $B$ or vacant to 
get the same result as above (detail not shown).
Therefore, we conclude that that block spin analysis supports the
claim that the NCIM constitutes a dynamic universality class that is
characterized by the continuously varying critical exponents.

\section{Summary and Discussion}\label{sec:5}
In this paper, we revisited the nonequilibrium chiral Ising model in one dimension 
using the block renormalization method introduced by Basu and Hinrichsen~\cite{Basu11},
mainly focusing on the maximal chiral model which was claimed to have continuously varying
exponents~\cite{Kim13}.
First introducing 5 different block spin states reflecting the symmetry of the system
as well as the property of having absorbing states, we calculated the asymptotic
value of block spin correlation functions which are expected to be universal. 
It turned out that (universal) ratio of block spin correlation functions
varies with a model parameter,
which along with the universality hypothesis supports the continuously varying nature of the MCM.

We also provided a microscopic theory about the scaling relation Eq.~\eqref{zd}
which associates the ratio of probability that a block with size $b$ is occupied by at least single particle
with the critical exponent $z\delta$. First, we clarified necessary conditions that a critical system
obeys Eq.~\eqref{zd}. Then, we found a relation between two-point correlation functions and 
the probability that exactly $r$ consecutive sites are empty using cluster mean field type approximation,
which is numerically found to be valid for the MCM. Finally, we estimated $z\delta$ using
Eq.~\eqref{zd} to find that $z \delta$ is continuously varying and is numerically consistent with
the previous numerical results, which again strongly supports that the continuously varying
exponents are the inherent feature of the MCM.

Although we neglected the symmetry due to chirality and only kept
the feature of having absorbing states when we define
$\rho_n$ in Sec.~\ref{sec:4}, we obtained the consistent scaling relation.
In this sense, the symmetry of the system is not crucial
in the block spin transformation of Basu-Hinrichsen formalism 
unlike the usual renormalization group theory.
The only important feature, at least for models with absorbing states,
is whether the block spin can capture the absorbing state properly.

\begin{acknowledgments}
MK acknowledges the financial support from the TJ Park Foundation. S-CP
acknowledges the support by the Basic Science Research Program through the
National Research Foundation of Korea~(NRF) funded by the Ministry of
Education, Science and Technology~(Grant No. 2011-0014680) and 
the hospitality of Asia Pacific Center for Theoretical Physics (APCTP).
This work is also supported by the the Basic Science
Research Program through the NRF Grant No.~2013R1A2A2A05006776.
\end{acknowledgments}

\appendix
\bibliographystyle{apsrev}
\bibliography{BR}
\end{document}